\documentstyle[12pt,aasms4]{article}
\def\be{\begin{equation}}
\def\ee{\end{equation}}
\def\ba{\begin{eqnarray}}
\def\ea{\end{eqnarray}}

\def\la{\mathrel{\mathpalette\fun <}}
\def\ga{\mathrel{\mathpalette\fun >}}
\def\fun#1#2{\lower3.6pt\vbox{\baselineskip0pt\lineskip.9pt
        \ialign{$\mathsurround=0pt#1\hfill##\hfil$\crcr#2\crcr\sim\crcr}}}

\slugcomment{To be submitted to ApJ}

\begin{document}
\null\vspace{-62pt}
\begin{flushright}
astro-ph/0101040\\
ApJ, 552, May 10 (2001)\\
{\it submitted on Aug.2, 2000}\\
\end{flushright}
\title{Measuring Time-Dependence of Dark Energy Density\\
 from Type Ia Supernova Data}

\vspace{0.1in}
\author{Yun Wang\footnote{current address:
Department of Physics \& Astronomy,
University of Oklahoma, Norman, OK 73019.
email: wang@mail.nhn.ou.edu}, \& Peter M. Garnavich}
\affil{{\it Center for Astrophysics\\
Dept. of Physics, 225 Nieuwland Science Hall\\
University of Notre Dame, Notre Dame, IN 46556-5670}}

\vspace{.4in}
\centerline{\bf Abstract}
\begin{quotation}

Observations of high redshift supernovae imply an accelerating Universe
which can only be explained by an unusual energy component  such as
vacuum energy or quintessence. To assess the ability of current and
future supernova data to constrain the properties of the dark energy,
we allow its density
to have arbitrary time-dependence, $\rho_X(z)$. This leads to an
equation of state for the dark energy,
$w_X(z)=p_X(z)/\rho_X(z)$, which is a free function of redshift $z$.
We find that current type Ia supernova (SNe Ia) data are consistent 
with a cosmological constant, with large uncertainties at $z\ga 0.5$.
We show that $\rho_X(z)/\rho_X(z=0)$ can be measured reasonably well
to about $z=1.5$ using type Ia supernova data from realistic future 
SN Ia pencil beam surveys,
provided that the weak energy condition (energy density of matter 
is nonnegative for any observer) is imposed.

While it is only possible to differentiate
between different models (say, quintessence and k-essence) at $z \la 1.5$
using realistic data,
the correct trend in the time-dependence of the dark energy density
can be clearly detected out to $z=2$, even in the presence of plausible 
systematic effects. This would allow us to determine
whether the dark energy is a cosmological constant, or some exotic form
of energy with a time-dependent density.

\end{quotation}


\section{Introduction}

Most of the present energy content of our universe is unknown (\cite{neta99}). 
Distance-redshift relations derived from cosmological standard candles
at redshifts between zero and a few are
the most sensitive probes of the equation of state of the universe,
which allows us to constrain the energy content of the universe.

Type Ia supernovae (SNe Ia) are our best candidates for cosmological 
standard candles.
They can be calibrated to have small dispersion in their intrinsic
luminosities (\cite{Phillips93,Riess95}). 
The data from two independent observational teams, 
the High-$z$ SN Search (Schmidt et al.) and
the Supernova Cosmology Project (Perlmutter et al.), 
seem to suggest that our universe has a significant vacuum energy content
(\cite{Garna98a,Riess98,Perl99}).

While a cosmological constant term (vacuum energy) in the Einstein equations
provides the simplest explanation of the current SN Ia data, other
forms of energy (quintessence, dark energy, etc) have also been 
studied (e.g. \cite{White98,Garna98b,Stein99,Efsta99,Ratra00,Josh00}) 
and are consistent with current data as well.
Since cosmology has matured into a phenomenological science at the
turn of the new millennium, observational data will dominate aesthetics
in the selection of cosmological models.

So far, most cosmologists have assumed time-independent equations of state,
i.e., power-law dark energy density (see \S 2),
in the context of constraining the energy content of the universe. 
In this paper, we allow the dark energy density
of the universe to be a free function of redshift, 
i.e., arbitrary equation of state. We
derive constraints on the dark energy density from current SN data,
and assess future prospects of measuring the dark energy density
using simulated data from
realistic future SN Ia surveys. 
We expect a model-independent measurement of the dark energy density as
a function of redshift to be very useful in our quest of the unknown
energy contents of the universe.

\section{Parametrization of the dark energy density}

In a smooth Friedmann-Robertson-Walker (FRW) universe,
the metric is given by $ds^2=dt^2-a^2(t)[dr^2/(1-kr^2)+r^2 (d\theta^2
+\sin^2\theta \,d\phi^2)]$, where $a(t)$ is the cosmic scale factor,
and $k$ is the global curvature parameter. The cosmological redshift $z$
is given by $1+z=1/a$.

To make model-independent measurements of the equation of state, we 
replace the vacuum energy density $\rho_{\Lambda}$ with 
$\rho_X=\rho_X^0 f(z)$ in the total matter density of the universe:
\ba
\rho(z) & = & \rho_m^0(1+z)^3 +\rho_k^0(1+z)^2 +\rho_X^0\, f(z)
\nonumber \\
 & = & \rho_c^0 \left[\Omega_m (1+z)^3 +\Omega_k(1+z)^2 + \Omega_X \,f(z)
\right],
\ea
where the superscript ``0'' indicates present values, $f(z=0)=1$, and
$\Omega_k=1-\Omega_m-\Omega_X=-k/H_0^2$.
If the unknown energy is due to a cosmological constant $\Lambda$,
$f(z)=1$. Clearly, the function $f(z)$ is a very good probe of
the nature of the unknown energy.

The comoving distance $r$ is given by (\cite{Weinberg72})
\be
\label{eq:r(z)}
r(z)=cH_0^{-1}\, \frac{S(\kappa \Gamma)}{\kappa}, \hskip 1cm
\kappa \equiv \left| \Omega_k \right|^{1/2},
\ee
\be
\Gamma(z;\Omega_m,\Omega_X, f)=\int_0^zdz' \frac{1}{E(z')},
\ee
\be
E(z) \equiv \left[ \Omega_m(1+z')^3+ \Omega_X\, f(z')+\Omega_k(1+z')^2
\right]^{1/2},
\ee
where
\ba
S(x)&=&\sinh(x), \hskip 2cm \Omega_k>0 \nonumber \\
&=&x, \hskip 3cm \Omega_k=0 \nonumber \\
&=&\sin(x), \hskip 2cm \Omega_k<0.
\ea
The angular diameter distance is given by $d_A(z)=r(z)/(1+z)$,
and the luminosity distance is given by $d_L(z)=(1+z)^2 d_A(z)$.

Einstein's equations in a FRW metric, together with the first law of 
thermodynamics give us
\be
\label{eq:energycons}
(1+z)\, \frac{ d\rho}{dz} = 3(\rho+p).
\ee
For unknown energy $\rho_X(z)$, we find
\ba
\rho_X(z) &=& \rho_X^0\, f(z) \nonumber \\
p_X(z) &=& \rho_X^0\, \left[ \frac{1}{3} (1+z) \, f'(z) - f(z) \right].
\ea
Now we can write the equation of state of the unknown energy as
\be
w(z) \equiv \frac{p_X(z)}{\rho_X(z)}=\frac{1}{3}(1+z) \frac{f'(z)}{f(z)} -1.
\ee
A constant equation of state corresponds to $f(z) \propto (1+z)^{\alpha}$,
where $\alpha$ is a constant. The values $\alpha=0$, $\alpha=3$, and $\alpha=4$ 
correspond to a cosmological constant, matter, and radiation
respectively. 

To obtain accelerated expansion, we need 
$\rho+3p<0$. Since $\rho_X+3p_X=\rho_X^0\,
\left[ (1+z) f'-2f\right]$, this implies
$\alpha<2$ for $f(z) \propto (1+z)^{\alpha}$.

The weak energy condition states that for all physically reasonable
classical matter, the energy density of matter as measured by any 
observer is nonnegative (\cite{Wald84}). For a perfect fluid, the
weak energy condition will be satisfied if and only if
\be
\rho+p \ge 0.
\ee
This leads to (see Eq.[\ref{eq:energycons}])
\be
f'(z) \ge 0.
\ee
The weak energy condition imposes strong constraints on the
jointly estimated cosmological parameters $\Omega_m$, $\Omega_{\Lambda}$,
and the dark energy density $f(z)$.

Note that the comoving distance $r(z)$ depends on the equation of state
of X through
\be
\Gamma(z) \equiv \int_0^z 
dz'\,\left[ \Omega_m(1+z')^3+ \Omega_X\, f(z')+\Omega_k(1+z')^2
\right]^{-1/2}=H_0 \int_0^r \frac{dr}{\sqrt{1-kr^2}},
\ee
Hence
\be
\label{eq:derivf}
\Gamma'(z) = \left[ \Omega_m(1+z)^3+ \Omega_X\, f(z)+\Omega_k(1+z)^2
\right]^{-1/2}=
\frac{H_0}{\left[1+\Omega_k H_0^2 r^2 \right]^{1/2}}\, \frac{dr}{dz}.
\ee
To measure $f(z)$ directly from data, we need to evaluate the derivative
of the distance $r(z)$ with respect to redshift $z$.

To avoid taking derivatives of noisy data, we can parametrize $f(z)$
with its values at $n$ equally spaced redshifts $z_i$, and assume that
$f(z)$ is given by linear interpolations at other values of $z$.
We write
\ba
\label{eq:f(z)}
&&f(z)=\left( \frac{z_i-z}{z_i-z_{i-1}} \right)\, f_{i-1}+
\left( \frac{z-z_{i-1}}{z_i-z_{i-1}} \right)\, f_i,
 \hskip 1cm z_{i-1} < z \leq z_i, \nonumber \\
&& z_0=0, \,\, z_{n}=z_{max}; \hskip 1cm f_0=1,\,\, f_n=f_{n-1}
\ea
where $f_i$ ($i=1,2,...,n-1)$ are independent variables
to be estimated from data. 

\section{Parameter estimation}

The measured distance modulus for a SN Ia is
\be
\mu_0^{(l)}= \mu_p^{(l)}+\epsilon^{(l)}
\ee
where $\mu_p^{(l)}$ is the theoretical prediction
\be
\label{eq:mu0p}
\mu_p^{(l)}= 5\,\log\left( \frac{ d_L(z_l)}{\mbox{Mpc}} \right)+25,
\ee
and $\epsilon^{(l)}$ is the uncertainty in the measurement, including
observational errors and intrinsic scatters in the SN Ia absolute
magnitudes. 

Denoting all the parameters to be fitted as {\bf s},
we can estimate {\bf s} using a $\chi^2$ statistic, with
(Riess et al. 1998)
\be
\chi^2(\mbox{\bf s})=
\sum_l \frac{ \left[ \mu^{(l)}_p(z_l| \mbox{\bf s})-
\mu_0^{(l)} \right]^2 }{\sigma_{\mu_0,l}^2 +\sigma_{mz,l}^2}
\equiv \sum_l \frac{ \left[ \mu^{(l)}_p(z_l| \mbox{\bf s})-
\mu_0^{(l)} \right]^2 }{\sigma_l^2 },
\ee
where $\sigma_{\mu_0}$ is the estimated measurement error of the distance
modulus, and $\sigma_{mz}$ is the dispersion in the distance modulus 
due to the dispersion in galaxy redshift, $\sigma_z$, due to
peculiar velocities and uncertainty in the galaxy redshift
(for the Perlmutter et al. data, the dispersion due to peculiar
velocities is included in $\sigma_{m_B^{eff}}$, i.e., $\sigma_{\mu_0}$).
Since
\be
\label{eq:sigmamz}
\sigma_{mz}=\frac{5}{\ln 10} \left( 
\frac{1}{d_L}\frac{\partial d_L}{\partial z} \right)\, \sigma_z,
\ee
$\sigma_{mz}$ depends on the parameters {\bf s}. 
The probability density function (PDF) for the parameters {\bf s} is
\be
p(\mbox{\bf s}) \propto \exp\left( - \frac{\chi^2}{2} \right).
\ee
The normalized PDF is obtained by dividing the above expression
by its sum over all possible values of the parameters {\bf s}.

In order to impose the weak energy condition $f'(z) \ge 0$, we
compute the PDFs on a $N$-dimensional grid for $N$ parameters.
The PDF of a given parameter $s_i$ is obtained by integrating over
all possible values of the other $N-1$ parameters.
To reduce the computation time, we can integrate over
the Hubble constant $H_0$ analytically, and define a modified
$\chi^2$ statistic, with
\be
\label{eq:chi2mod}
\tilde{\chi}^2 \equiv \chi_*^2 - \frac{C_1}{C_2} \left( C_1+ 
\frac{2}{5}\,\ln 10 \right),
\ee
where
\ba
\chi_*^2 &\equiv& \sum_l \frac{1}{\sigma_l^2} \left( \mu_{p}^{*(l)}-
\mu_{0}^{(l)} \right)^2, \nonumber \\
C_1 &\equiv& \sum_l \frac{1}{\sigma_l^2} \left( \mu_{p}^{*(l)}-
\mu_{0}^{(l)} \right), \nonumber \\
C_2 &\equiv& \sum_l \frac{1}{\sigma_l^2},
\ea
where
\be
\mu_p^* \equiv \mu_p(h=h^*)=42.384-5\log h^*+ 5\log \left[H_0 r(1+z)\right].
\ee
We take $h^*=0.65$. Our results are independent of the choice of $h^*$.

After experimenting with a number of different techniques, we 
developed an adaptive iteration method of 
estimating $f(z)$ based on the requirement
that $f'(z)\ge 0$ (i.e., the weak energy condition is satisfied). Starting 
with the initial guess of $f(z)=f(z=0)=1$ (a cosmological constant), we 
iteratively build up $f(z)$
as parametrized by Eq.(\ref{eq:f(z)}) while minimizing $\chi^2$.

\section{Constraints of $\rho_X(z)$ from current SNe Ia data}

Wang (2000b) has combined the data of the High-$z$ SN Search team 
(\cite{Schmidt98,Garna98a,Riess98})
and the Supernova Cosmology Project (\cite{Perl99}), yielding
a total of 92 SNe Ia. Using the method described in the previous
section, we estimate $\Omega_m$, $\Omega_X$, and $f(z)$ simultaneously
by minimizing the modified $\chi^2$ (see Eq.(\ref{eq:chi2mod})).

Fig.1 shows the dimensionless dark energy density $f(z)$ (as
parametrized by Eq.(\ref{eq:f(z)})) measured from this set of
92 SNe Ia. It is consistent with $f(z)=1$ (a cosmological constant),
with large uncertainty beyond $z\ga 0.5$. Note that none of the error
bars extend beneath $f(z)=1$, because we have
imposed the weak energy condition, i.e., $f'(z) \ge 0$, which implies
that $f(z) \ge f(z=0)=1$.
The estimated values of $\Omega_m=.3$ (0, .9); $\Omega_X=1.7$ (.4, 2.2)
are consistent with previous results (\cite{Garna98a,Riess98,Perl99,Wang00b}).
The errors are estimated from the ranges of parameters for which
$\chi^2=\chi^2_{min}+1$.

Wang (2000b) found that when fit to a model with a cosmological constant
as the dark energy, flux-averaging changes the best fit model to this data 
set of 92 SNe Ia. Without flux-averaging, the best fit model
is a closed universe with $\Omega_m=0.7\pm0.4$, and a vacuum energy 
density fraction of $\Omega_{\Lambda}=1.2\pm0.5$, consistent with
the estimated values in Fig.1.
The flux-averaged data yield 
$\Omega_m=0.3\pm0.6$, and $\Omega_{\Lambda}=0.7\pm0.7$. 
This difference may have resulted from the data containing large redshift 
dependent uncertainties, which would have caused the results from the
unbinned data to be biased. 
The effect of flux-averaging on the best fit model assuming an arbitrary
dimensionless dark energy density $f(z)$ will be studied elsewhere.

Future cosmic microwave background (CMB) space missions 
MAP (\cite{Bennett97}) and Planck (\cite{DeZotti99}), together with
the galaxy redshift surveys SDSS (\cite{Gunn99}) and 2df (\cite{Dalton00}),
will give us exquisitely accurate measurements of the geometry of the
universe and the matter density in the universe 
(\cite{Eisen99,Mike99,Wang99a}). 
SN data can provide the unique probe on the nature of dark energy
by allowing us to measure how the dark energy density varies
with time.

Current cosmic microwave background (CMB) anisotropy measurements seem to 
indicate that we live in a flat universe (\cite{deBernardis00,Balbi00}).
Cluster abundances strongly suggest a low matter density 
universe (\cite{neta95,Carlberg96,neta98}). 
$\Omega_m=0.3$ and $\Omega_{\Lambda}=0.7$
is the best fit model to current observational data.
We will use $\Omega_m=0.3$ and $\Omega_X=0.7$ for our simulated data
in the rest of this paper.

\section{Measuring $\rho_X(z)$ from future SNe Ia data}

A large number of SNe Ia at $z\ga 1$ is critical in
resolving the important systematic uncertainties 
of SNe Ia as cosmological standard candles, such as
dust (\cite{Aguirre99}), gravitational lensing 
(\cite{Kantow95,Wamb97,Holz98,Metca99,Wang99b,Barber00}), 
and luminosity evolution (\cite{Drell99,Riess99,Wang00b}),
and in making SNe Ia useful probes of the dark energy
content of the universe. The most efficient method of obtaining
a large number of SNe Ia at $z>1$ is conducting a supernova pencil
beam survey on a dedicated large aperture telescope with a square 
degree field of view (\cite{Wang00a}).

To study how well SN data can probe the dark energy density, let us consider 
two hypothetical dimensionless dark energy densities 
$f_q(z)$ and $f_k(z)$, given by
\ba
\label{eq:f(z)1}
f_q(z)&=& \frac{ e^{1.5z} }{(1+z)^{1.5}}, \hskip 1cm w_q(z)= -1+0.5z \nonumber \\
f_k(z) &=& \exp[0.9(1-e^{-z})], \hskip 1cm w_k(z)= 0.3 (1+z) e^{-z} -1 
\ea
We have chosen $f_q(z)$ and $f_k(z)$ to represent quintessence 
($dw_q/dz >0$) and k-essence ($dw_k/dz <0$) models respectively
(\cite{Caldwell98,Armenda00}).
Note that $f_q(z)$ and $f_k(z)$ satisfy the weak energy condition
$f'(z)\ge 0$; they give an accelerating universe for $z\la 1.33$ 
and $z\la 2$ respectively.

A feasible SN pencil beam survey (either from 
ground\footnote{The SNe Ia at $z \ga 1.5$ will
likely require follow up spectroscopy from space.}
or from space), 
with a square degree field of view and for an effective observational 
period of one year, can yield almost 2000 SNe Ia out to $z=2$ 
(\cite{Wang00a}).
Let us combine the data from the SN pencil beam survey 
with SN data at smaller redshifts, so that
there are a minimum of 50 SNe Ia per 0.1 redshift interval
at any redshift.
This yields a total of 1966 SNe Ia for the quintessence and
1898 SNe Ia for the k-essence model,
up to $z=2$ and for $\Omega_m=0.3$, $\Omega_X=0.7$.
We simulate the data by placing perfect
standard candles at random redshifts, with the total number per 0.1
redshift interval given as described above. Then we add intrinsic 
and observational dispersions which are Gaussian with zero mean and
a variance of 0.20 magnitudes. 
A systematic shift in $\mu_0$ of $dm_{sys}\,z$ magnitudes is
also added to mimic possible systematic errors as one 
goes to larger redshifts.

Realistic SN Ia data should contain gravitational lensing noise.
Wang (2000a,b) has shown that flux averaging is important in reducing
the bias due to lensing. The effect of gravitational lensing and
of flux averaging in the context of a general equation of state will
be investigated elsewhere.

Fig.2 illustrates how well we can recover the dark energy densities
$f_q(z)$ and $f_k(z)$ in the absence of lensing
noise and systematic shifts, when we apply 
our adaptive iteration method to 100 random data sets with 
a realistic dispersion of 0.2 magnitudes. 
To study the dependence of our results on the parametrization
of $f(z)$ [see Eq.(\ref{eq:f(z)})], we show results for 
(a) $n=10$, and (b) $n=6$.
The thick and thin solid lines are the assumed true 
$f_q(z)$ and $f_k(z)$ respectively.
We have assumed that we know $\Omega_m+\Omega_X=1$.
The mean and 1-$\sigma$ errors of $f(z)$ and $\Omega_m$ 
are estimated from averaging over the 100 random samples.
The error of estimated $\Omega_m$ reflects the resolution of the program.
For a given data set, the recovered $f(z)$ and $\Omega_m$ are expected
to fall within the errors with 68\% probability.
Clearly, the quintessence and k-essence models can be
differentiated marginally for $z\la 1.5$. For $z \ga 1.5$,
the errors increase significantly, while the estimates become 
more biased, making it impossible to differentiate between
the two models. However, the correct trend in the time
variation of the dark energy density can be clearly
detected out to $z=2$.

Fig.2 shows that the parametrization of $f(z)$ with $n=10$ yields
less biased estimates than $n=6$. This is as expected, since 
$f(z)$ is more accurately parametrized as one
increases $n$. However, the errors in the estimates increase
with $n$ as well. One must choose an optimal $n$ such that $f(z)$ 
is adequately parametrized, while the errors on the estimated
amplitudes of $f(z)$ are not too large to be useful.
We've experimented with $n>10$ parametrizations of $f(z)$,
and found that $n=10$ is a good choice.

The biased estimates of the dark energy density $f_q(z)$ and $f_k(z)$
in Fig.2 is mainly due to the bias in the estimate of $\Omega_m$.
Fig.3 shows $f_q(z)$ and $f_k(z)$ estimated assuming that we know
$\Omega_m=0.3$, with the same line types as in Fig.2.
Even with this ideal assumption, it is only
possible to marginally differentiate between the two models.

Fig.4 shows the effect of adding a systematic shift of $dm_{sys}\,z$
to 100 random data sets with a realistic dispersion of 0.2 magnitudes.
The line types are the same as in Fig.2. 
We have added a systematic shift in $\mu_0$ of
(a) $0.01\,z$ magnitudes; (b) $0.05\,z$ magnitudes.
Clearly, systematic shifts can significantly increase the
bias in the estimate of $\Omega_m$ and the estimates of
$f(z)$ for $z\ga 1.2$, while having little effect
on the estimates of $f(z)$ at $z\la 1.2$.
Systematic errors as a function of $z$ may arise from intrinsic
properties of the supernovae varying with $z$ such as progenitor
evolution or dust characteristics changing with metalicity of the Universe.
Accuracy may also be limited by the observations themselves in the
case of k-corrections or selection biases. The estimated errors from
these sources in the current surveys are between 2\% and 
10\% (\cite{Schmidt98}). 
It is perhaps not realistic to expect to measure
both $\Omega_m$ and $f(z)$ to high redshift accurately from SN Ia data alone.

It is clear from Fig.4 that even in the presence of plausible
systematic effects, we can expect to measure the time-variation in
the dark energy density $f(z)$ with reasonable accuracy to
a redshift of about 1.2. The bias and the errors in the estimated $f(z)$
increase substantially beyond $z=1.2$.

We only applied our adaptive iteration method to 100 random samples,
and with a resolution of $\Delta\Omega_m=0.02$,
because this method takes several hours per sample on a fast
Sun work station in finding
the best fit $f(z)$ and $\Omega_m$. It is work planned for the future
to improve this promising method
for application to much larger number of random samples, as well as
adding $\Omega_X$ as an estimated parameter.

\section{Implication for dark energy models}

Recently, there has been a great deal of activity in
exploring the possibilities of the existence of exotic dark energy 
(\cite{Peebles88,Frieman95,Caldwell98,Sahni00})
in the universe. While the present
observational data are consistent with the dark energy being a cosmological
constant, they do not rule out alternatives in the form 
of various scalar fields.

It is important that we measure the time dependence of the dark energy density.
If the dark energy density is measured to be constant in time
within reasonable uncertainties, a cosmological constant should 
be favored, and more theoretical efforts should be directed toward
the derivation of a cosmological constant from first principles.
At the very least, this places strong constraint on the classes of
scalar-field models for the dark energy.
On the other hand, if the time-dependence of the dark energy density is
established by the observational data, we would come to the exciting
discovery of new physics in the universe.

We have studied two dark energy models $f_q(z)$ and $f_k(z)$
(see Eq.(\ref{eq:f(z)1})),
representing two general classes of dark energy models,
quintessence (\cite{Caldwell98}) and k-essence (\cite{Armenda00}).
This allows us to examine how well different models can
be differentiated by realistic data, as well as the robust
determination of the time-dependence of the dark energy density.

Realistic future SN data, as described in the previous section, 
has the potential of determining the time-dependence of the dark energy density.
This will clearly have a dramatic impact on models of the dark energy.

\section{Conclusions}

To access the prospects of measuring the time variation in the
equation of state, 
we have developed a promising adaptive iteration method that is
powerful in extracting the dark energy density $f(z)=\rho_X(z)/\rho_X(z=0)$ 
from realistic data. This method is based on the requirement 
that the weak energy condition (energy density of matter 
is nonnegative for any observer) is satisfied.

We have found that current type Ia supernova (SNe Ia) data are consistent 
with a cosmological constant, with large uncertainties at $z\ga 0.5$.
We show that $\Omega_m$ (assuming a flat universe) and
the dimensionless dark energy density
$f(z)=\rho_X(z)/\rho_X(z=0)$ can be measured reasonably well
to about $z=1.5$ using type Ia supernova data from realistic future 
SN Ia pencil beam surveys, 
provided that the weak energy condition (energy density of matter 
is nonnegative for any observer) is imposed.
For $z \ga 1.5$,
the errors increase significantly, while the estimates become 
more biased, making it impossible to differentiate between
different models (say, quintessence and k-essence).
However, the correct trend in the time-dependence of the dark energy density
can be clearly detected out to $z=2$, even in the presence of 
plausible systematic effects. This would allow us to determine
whether the dark energy is vacuum energy, or some exotic form
of energy with a time-dependent density.

The simulated data we used are for a SN pencil beam survey 
(either from ground or from space) with a square degree field of 
view and for an effective observational 
period of one year (\cite{Wang00a}), combined with SN data at 
smaller redshifts, so that
there are a minimum of 50 SNe Ia per 0.1 redshift interval
at any redshift. Although the dispersion of 0.20 magnitudes 
(intrinsic plus observational) assumed in our simulated data
is appropriate for ground-based surveys,
we expect our results to apply qualitatively to space based 
SN pencil beam surveys as well,
because our assumed dispersion of 0.20 magnitudes 
is dominated by intrinsic dispersion (about 0.17 magnitudes).

At the completion of this lengthy numerical study of the feasibility
of measuring the time-dependence of the dark energy density from
realistic SN Ia data, we became aware of the recent paper by
Maor, Brustein, \& Steinhardt (2000). They claimed that 
distance-redshift relations derived from SNe Ia and similar
classical measures are poor methods for resolving the time-dependence
or measuring the amplitude of the equation of state $w_X(z)=p_X(z)/\rho_X(z)$,
and consequently no useful information can be obtained about the
future fate of the universe. 
Our work has confirmed the difficulty of measuring the properties
of the dark energy from realistic SN Ia data. 
However, we have found their assessment to
be overly pessimistic. Their work indicates that it is impossible
to tell whether the equation of state $w_X(z)$ varies in time
(see also, \cite{Barger01}), 
but knowing that the dark energy density $\rho_X(z)$ varies in time 
is sufficient to rule out vacuum energy
as dark energy, thus giving support to exotic dark energy models.
Our work has shown that one can indeed clearly detect the
time-dependence of $\rho_X(z)$ using realistic future SN Ia data.

The main problem reported by Moar et al. was the ``smearing'' effect
of the multi-integral relation between the luminosity distance $d_L(z)$
and the equation of state $w_X(z)$. Instead of $d_L(z)$ and $w_X(z)$,
our analysis uses the time derivative of the comoving distance $r'(z)$ and 
the dimensionless dark energy density $f(z)$. Thus our results
are less affected by the smearing effect. Our method will be refined
and made more efficient, and should become quite useful in analyzing
future SN data.

In view of our results,
it is important that reasonable yet substantial observational
efforts are devoted to future SN Ia surveys, for example,
a dedicated 4m telescope for a SN pencil beam survey (\cite{Wang00a}),
combined with surveys of nearby SNe Ia. The total cost of such surveys
would be modest compared to the great scientific return,
the determination of the systematic uncertainties of SNe Ia
as cosmological standard candles, and the measurement of
the time-dependence in the dark energy density of the universe
to constrain fundamental physics.

\acknowledgements{\centerline{\bf Acknowledgements}}

It is a pleasure for us to thank Grant Mathews for helpful discussions,
and the referee for useful suggestions.
PMG acknowledges support from NASA LTSA grant NAG-9364.



\clearpage
\setcounter{figure}{0}

\figcaption[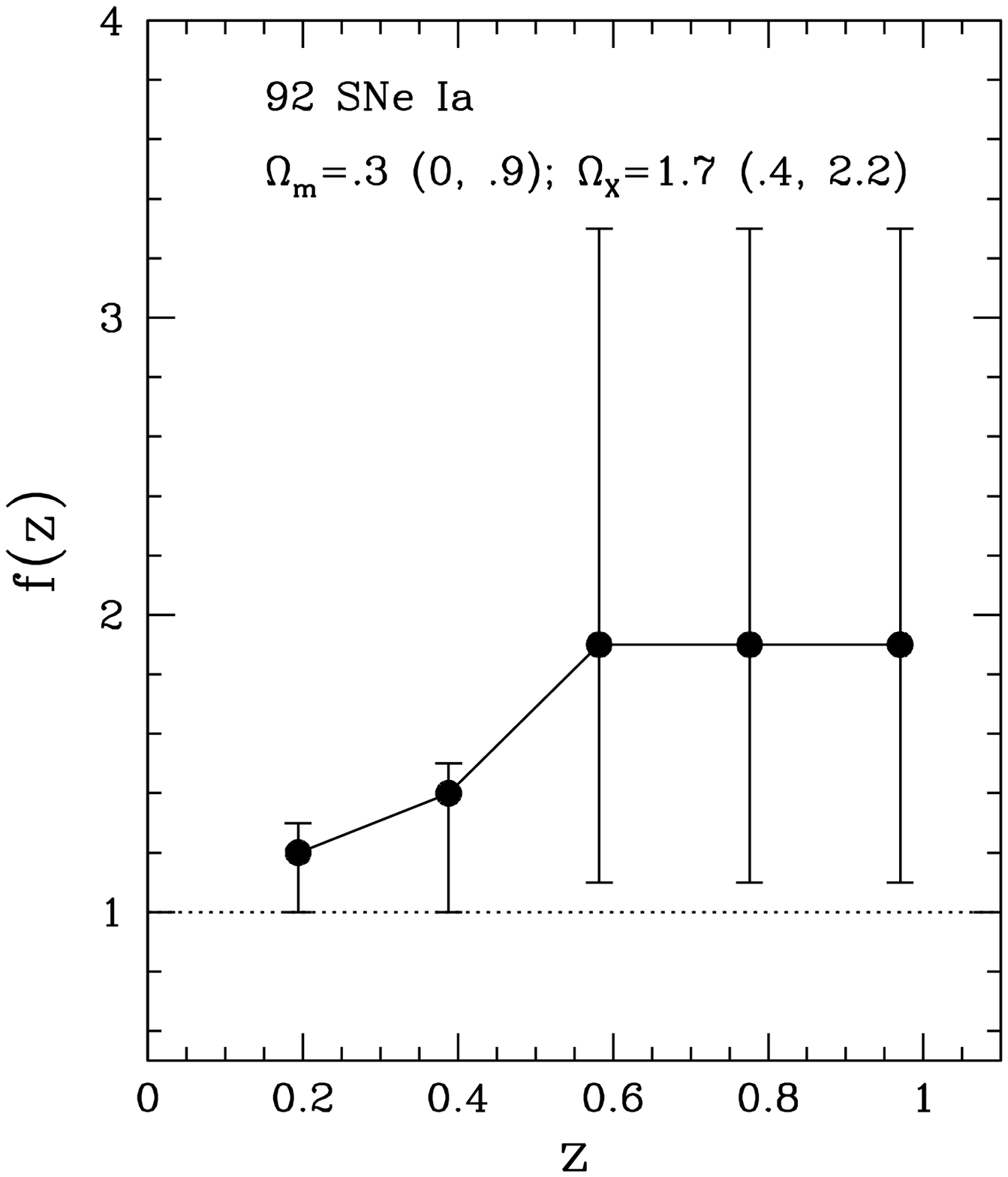]
{The dimensionless dark energy density $f(z)$ (as
parametrized by Eq.(\ref{eq:f(z)})) measured from the current combined 
data of 92 SNe Ia.}

\figcaption[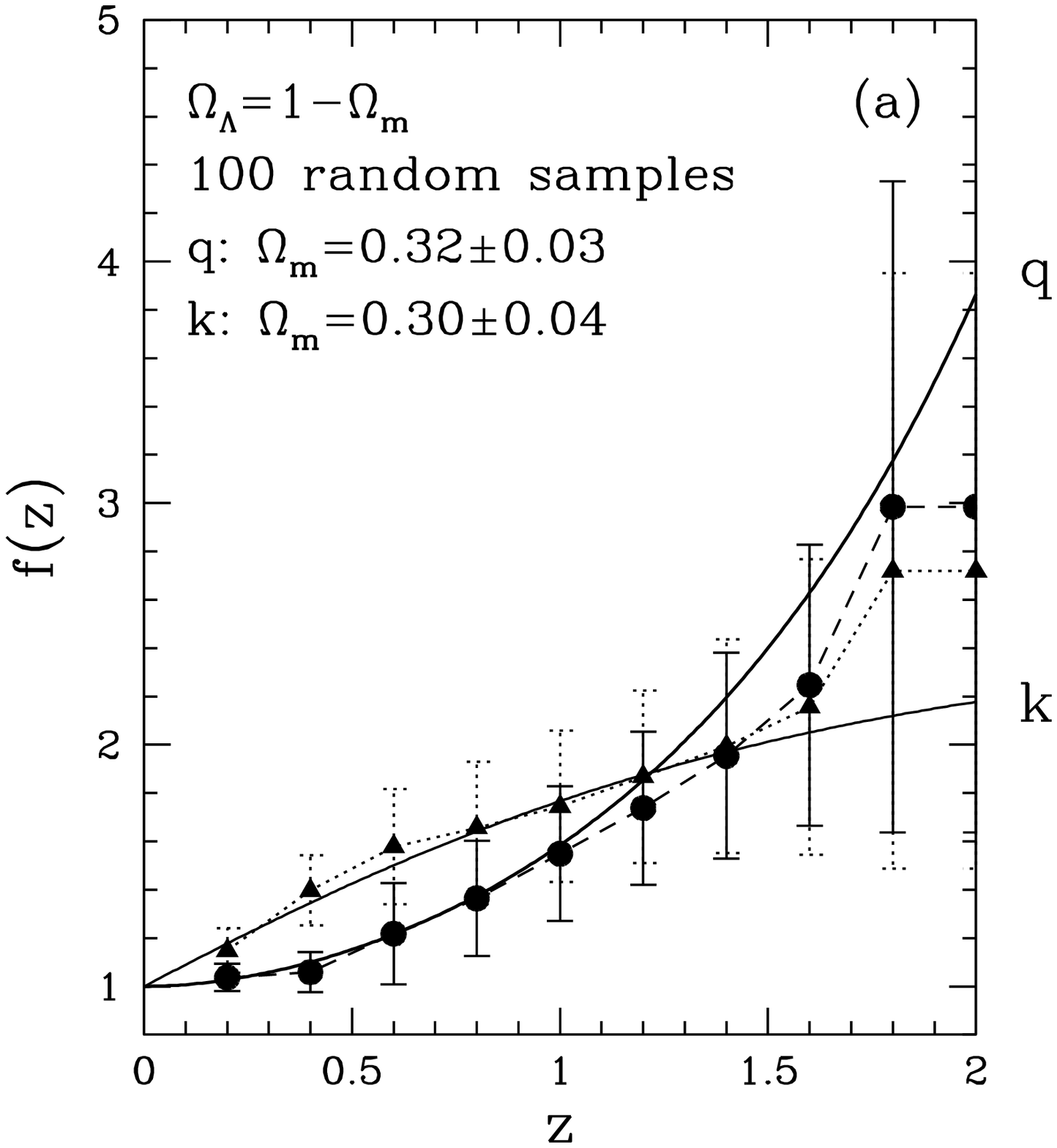]
{The measured dark energy density $f(z)$ [see Eq.(\ref{eq:f(z)})]
in the absence of lensing
noise and systematic shifts, when we apply 
our adaptive iteration method to 100 random data sets with 
a realistic dispersion of 0.2 magnitudes, 
for (a) $n=10$, and (b) $n=6$.
The thick and thin solid lines are the assumed true 
$f_q(z)$ and $f_k(z)$ respectively.
The circles and triangles are the estimated values of 
$f_q(z)$ and $f_k(z)$ respectively.}

\figcaption[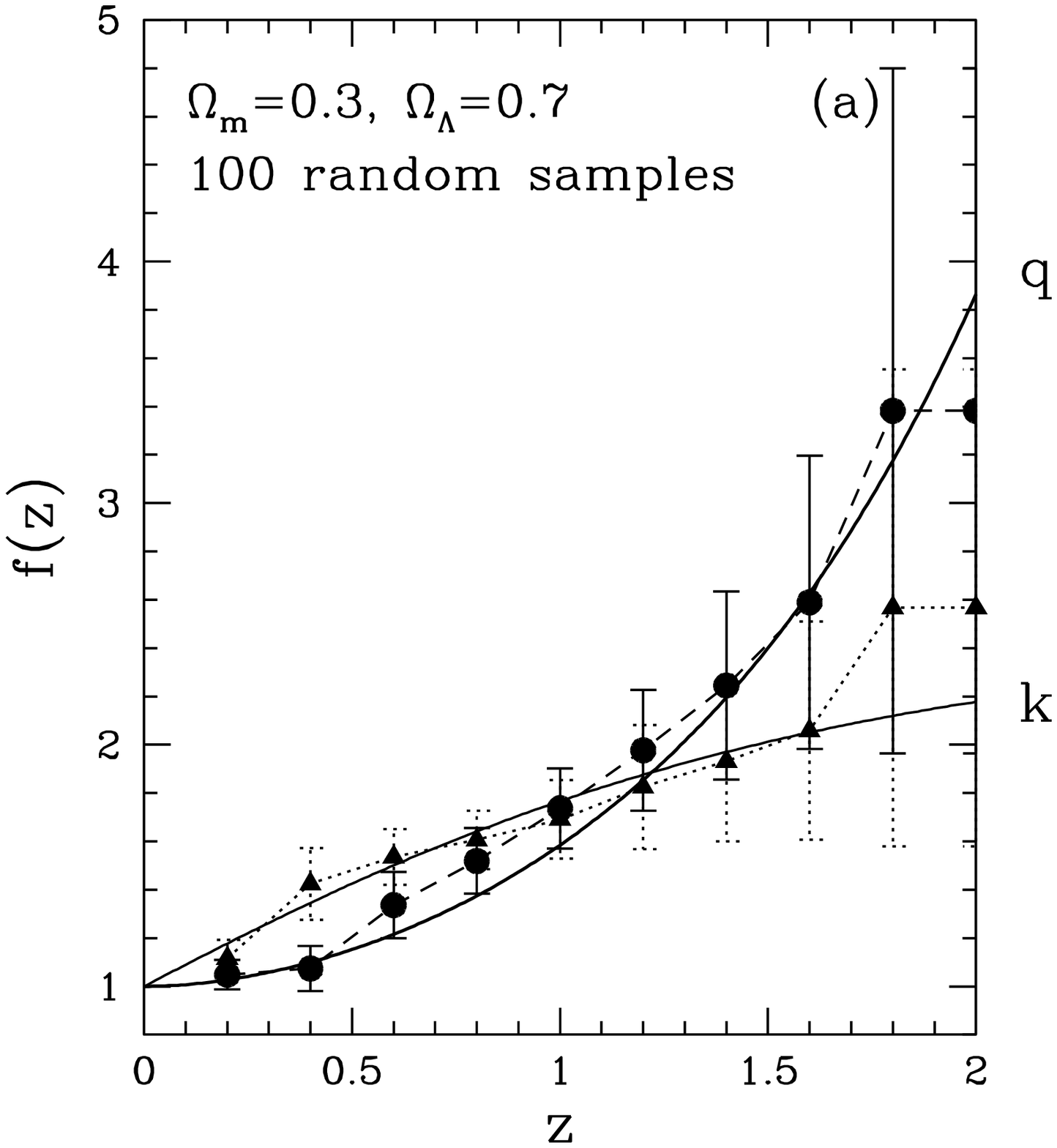]
{The dark energy densities $f_q(z)$ and $f_k(z)$ estimated assuming that 
we know $\Omega_m=0.3$, with the same line types as in Fig.2.}

\figcaption[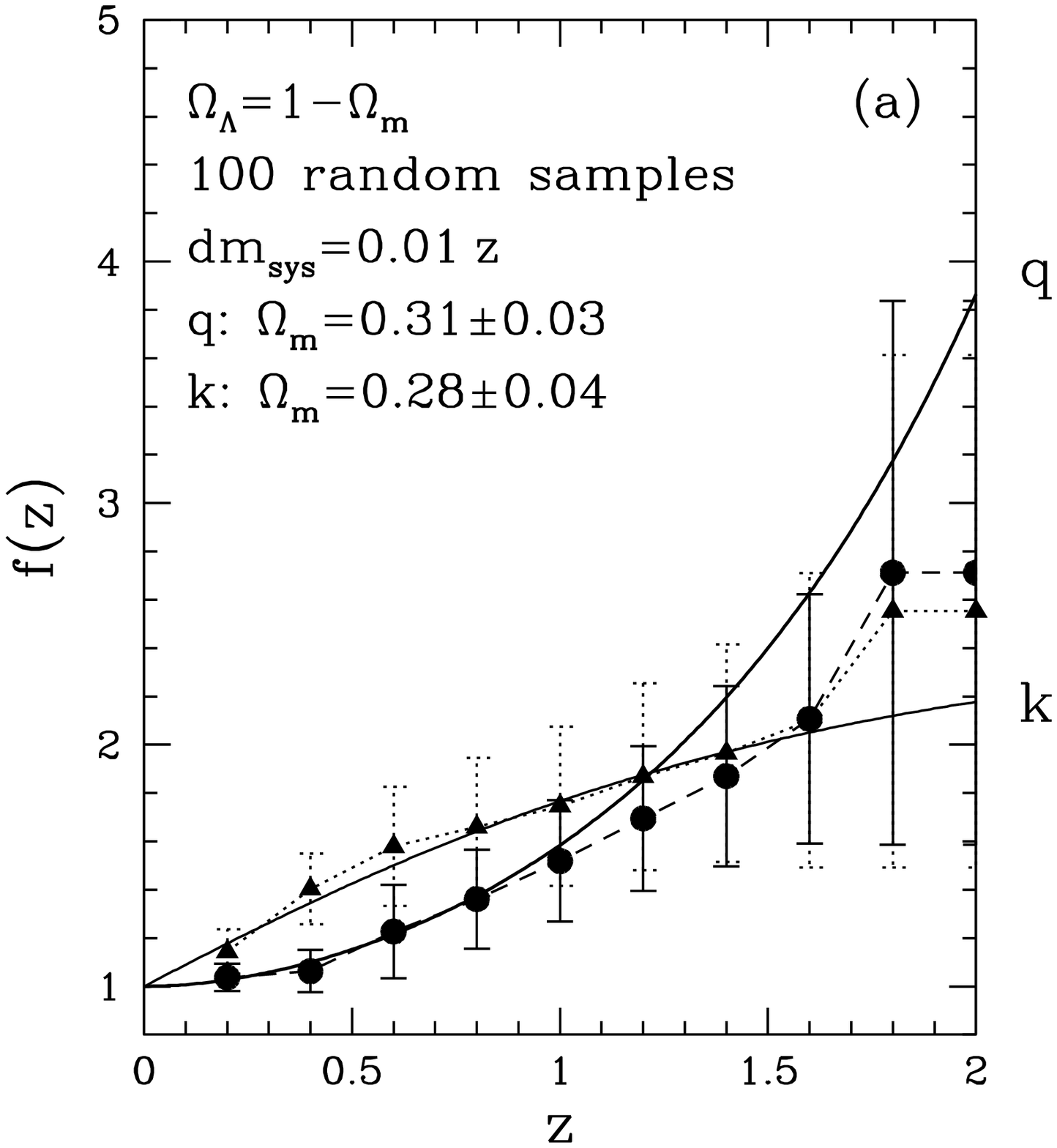]
{The effect of adding a systematic shift of $dm_{sys}\,z$
to 100 random data sets with a realistic dispersion of 0.2 magnitudes.
The line types are the same as in Fig.2.
We have added a systematic shift in $\mu_0$ of
(a) $0.01\,z$ magnitudes; (b) $0.05\,z$ magnitudes.}


\clearpage

\setcounter{figure}{0}
\plotone{fig1.eps}
\figcaption[fig1.eps]
{The dimensionless dark energy density $f(z)$ (as
parametrized by Eq.(\ref{eq:f(z)})) measured from the current combined 
data of 92 SNe Ia.}

\plotone{fig2a.eps}
\figcaption[fig2a.eps]
{The measured dark energy density $f(z)$ [see Eq.(\ref{eq:f(z)})]
in the absence of lensing
noise and systematic shifts, when we apply 
our adaptive iteration method to 100 random data sets with 
a realistic dispersion of 0.2 magnitudes. 
The thick and thin solid lines are the assumed true 
$f_q(z)$ and $f_k(z)$ respectively.
The circles and triangles are the estimated values of 
$f_q(z)$ and $f_k(z)$ respectively.
(a) $n=10$.}

\setcounter{figure}{1}
\plotone{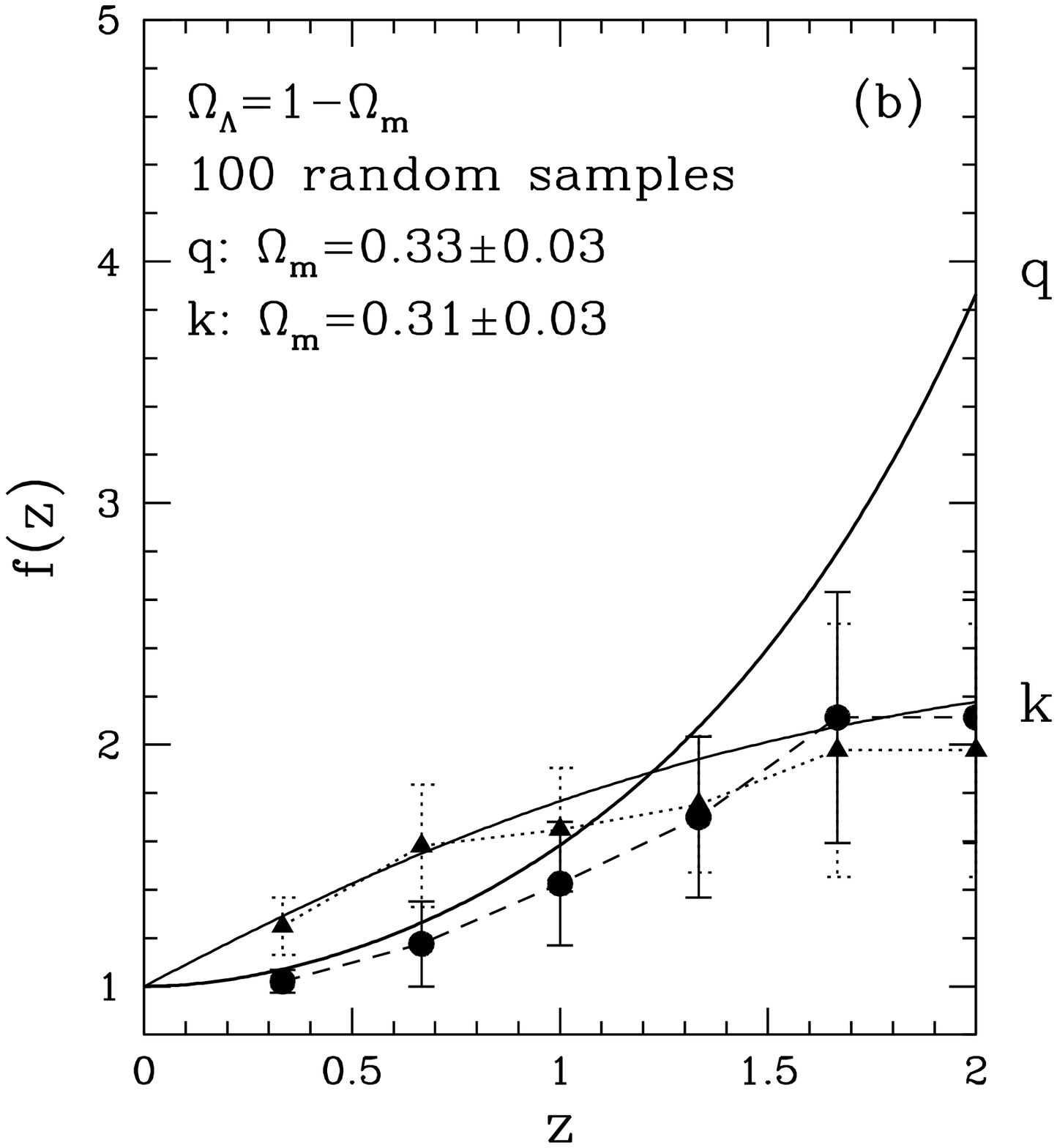}
\figcaption[fig2b.eps]
{(b) $n=6$.}

\plotone{fig3a.eps}
\figcaption[fig3a.eps]
{The dark energy densities $f_q(z)$ and $f_k(z)$ estimated assuming that 
we know $\Omega_m=0.3$, with the same line types as in Fig.2.
(a) $n=10$.}

\setcounter{figure}{2}
\plotone{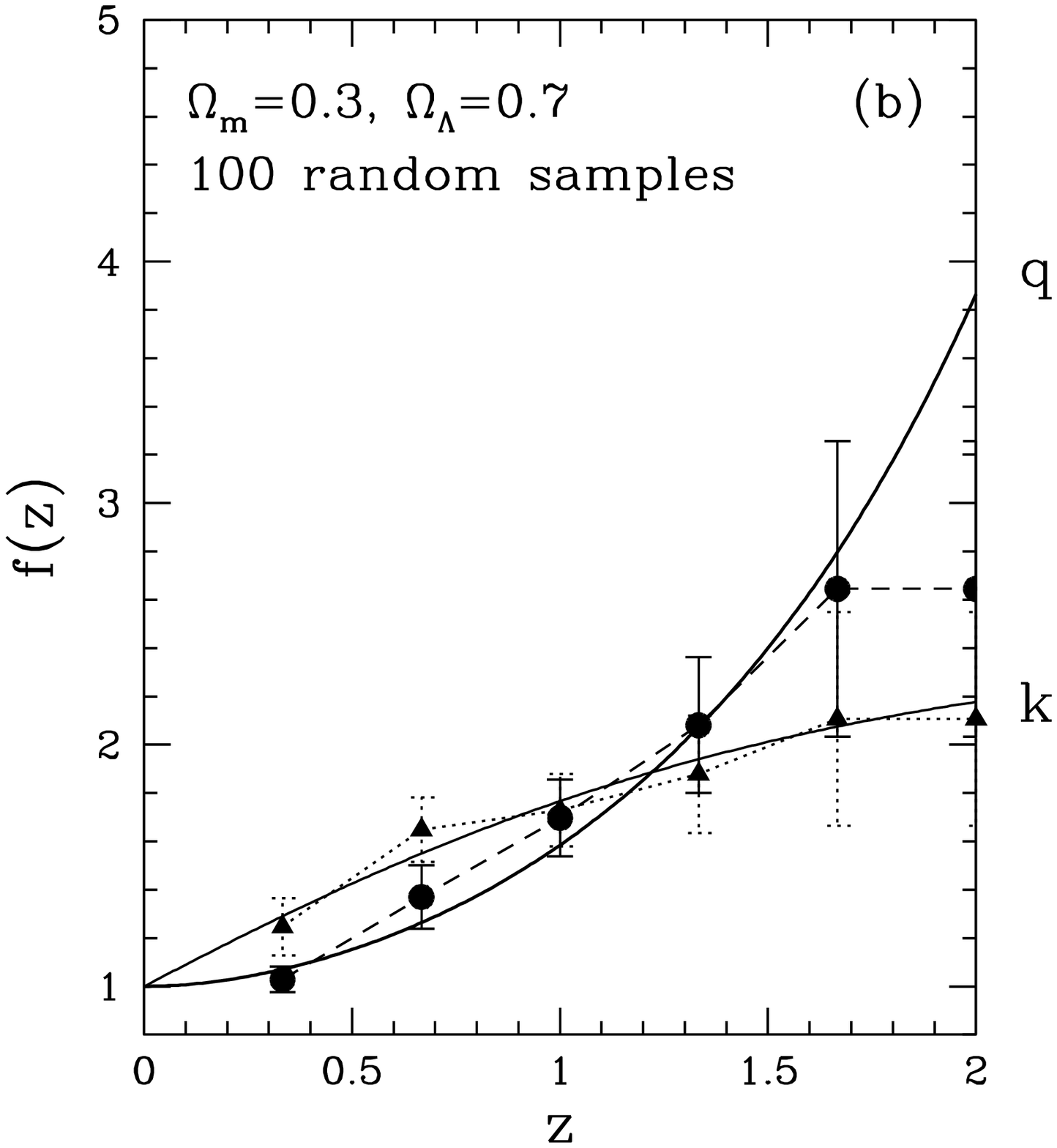}
\figcaption[fig3b.eps]
{(b) $n=6$.}

\plotone{fig4a.eps}
\figcaption[fig4a.eps]
{The effect of adding a systematic shift of $dm_{sys}\,z$
to 100 random data sets with a realistic dispersion of 0.2 magnitudes.
The line types are the same as in Fig.2.
(a) We have added a systematic shift in $\mu_0$ of $0.01\,z$ magnitudes.}

\setcounter{figure}{3}
\plotone{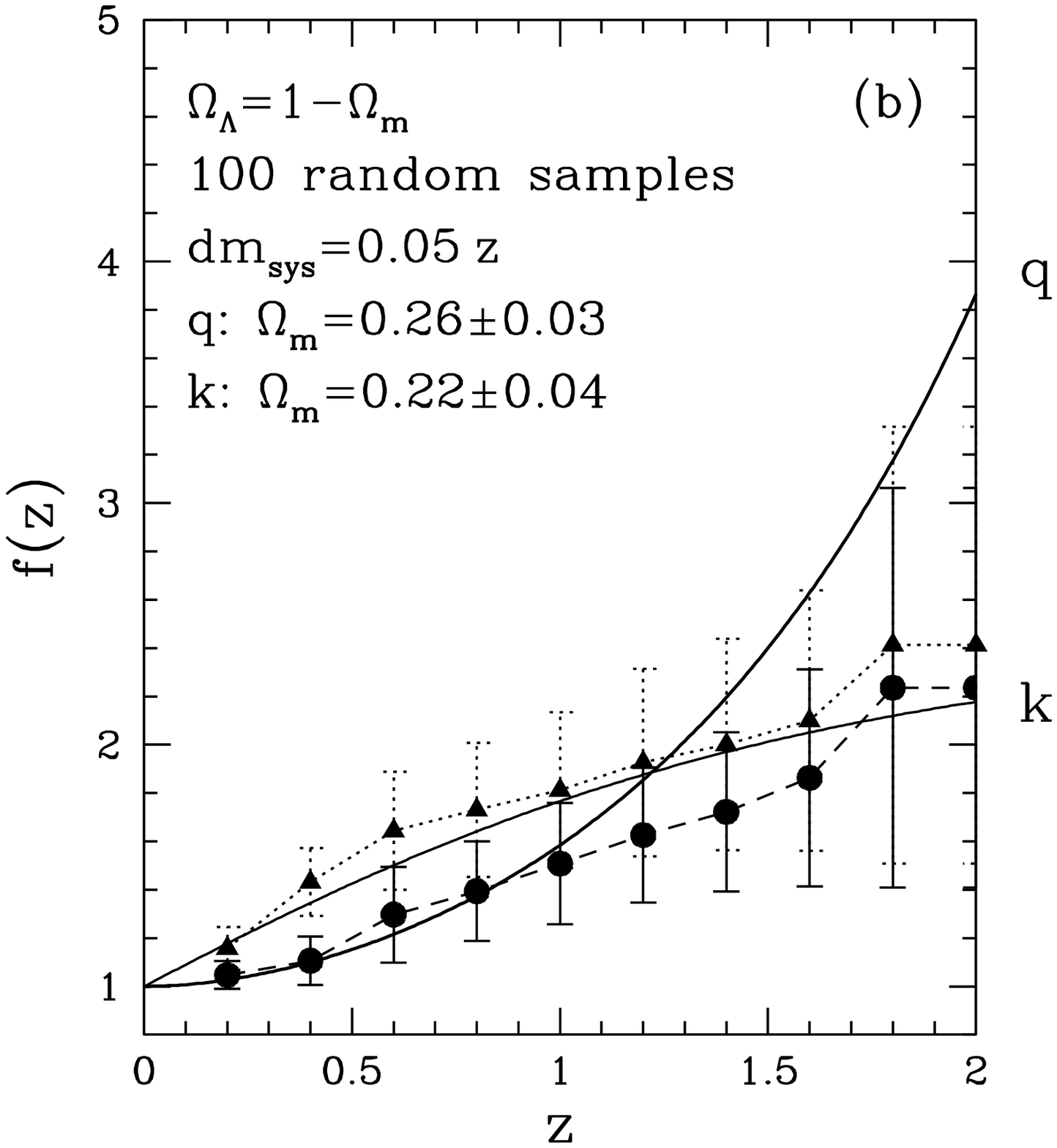}
\figcaption[fig4b.eps]
{(b) We have added a systematic shift in $\mu_0$ of $0.05\,z$ magnitudes.}

\end{document}